\documentstyle{article}
\title{Resummation of the perturbation series
for an interaction of a scalar field with a quantized metric }
\author{ Z. Haba\\Institute of Theoretical Physics, University of Wroclaw,
\\50-204 Wroclaw, Plac Maxa Borna 9, Poland\\e-mail:zhab@ift.uni.wroc.pl}
\date{}
\begin{document}
\maketitle \baselineskip 24pt
\begin{abstract}
We consider a scalar field interacting with a quantized metric
varying only on a submanifold (e.g. a scalar field interacting
with a quantized gravitational wave).
 We explicitly sum up
the perturbation series for the time-ordered vacuum expectation values
 of the scalar field. As a result we obtain
a modified non-canonical short distance behavior.

\end{abstract}
\section{ Introduction}
We consider a model of a scalar field interacting with a quantized
metric. In order to simplify the model we assume that the metric
does not depend on all the coordinates. There is a physical model
for such a metric tensor:  the field of a gravitational wave.
Hence, the model would describe an interaction of a scalar
particle with the graviton . We have obtained in \cite{jmp} a path
integral representation of the scalar field correlation functions.
Then, we assumed an exact scale invariance of the quantum
gravitational field. In this paper we insert a flat background
metric and consider an expansion around this metric. We show that
the path integral formula can be considered as a resummation of
the conventional perturbation expansion. Then, the assumption that
the perturbation around the flat metric is scale invariant leads
to the same conclusions as in \cite{jmp}:  the singular quantum
metric substantially changes the short distance behavior of the
scalar field (in particular, it can make the scalar field more
regular).

\section{The conventional perturbation expansion}
We consider the Lagrangian for a complex scalar
field in $D$-dimensions interacting with gravity
\begin{equation}
{\cal L}={\cal L}(g)
+g^{AB}\partial_{A}\overline{\phi}\partial_{B}\phi
+ M^{2}\overline{\phi}\phi
\end{equation}
 here ${\cal L}(g)$ is a gravitational Lagrangian which we do not
 specify  .
The metric is assumed to be a perturbation of the flat one
\begin{equation}
g^{\mu\nu}=\eta^{\mu\nu}+h^{\mu\nu}
\end{equation}
where $\mu,\nu=1,...,D-d$, $h^{\mu\nu}$ is varying only on a
$d$-dimensional submanifold and
\begin{equation}
g^{ik}=\eta^{ik}
\end{equation}
for $i,k=D-d+1,...,D$, here $\eta^{AB}=\pm\delta^{AB}$, but we do
not specify the signature.
 We split the coordinates as
  $x=({\bf X},{\bf x})$ with ${\bf x}\in R^{d}$.

As an example we could consider in $D=4$ the metric corresponding
to a gravitational wave  moving in the
$x_{3}$ direction. In such a case the metric  is
\begin{displaymath}
ds^{2}=-(dx_{4})^{2}+(dx_{3} )^{2}+g^{\mu\nu}dx_{\mu}dx_{\nu}
\end{displaymath}
where $g^{\mu\nu}(x_{3}-x_{4})$ is an exact solution of the
Einstein equations ,see e.g.\cite{inv} (this case would correspond
to $d=1$). More general solutions are known describing a
scattering of gravitational waves and depending on both
$x_{4}-x_{3}$ and $x_{4}+x_{3}$ (hence, $d=2$) ( for higher
dimensions see \cite{pen}\cite{tse}). The assumption of a weak
gravitational field (2) is a realistic approximation. In such a
case it can be shown that any solution of linearized Einstein
equations by means of a coordinate transformation  can be brought
to the form where all components of $h^{\mu\nu}$ are equal to zero
except of $\mu,\nu=1,2$ . These components describe the transverse
polarization states of
 the spin 2 field.

Without a self-interaction the $\phi\overline{\phi}$ correlation
function
 averaged over the metric is
\begin{equation}
\langle ({\cal A}+\frac{1}{2}M^{2})^{-1}(x,y)\rangle
\end{equation}
where
\begin{equation}
\begin{array}{l}
-{\cal A}= \frac{1}{2}\Box_{D-d}+\frac{1}{2}
\sum_{\mu=1,\nu=1}^{D-d}h^{\mu\nu}({\bf
x})\partial_{\mu}\partial_{\nu}+
\frac{1}{2}\sum_{k=D-d+1}^{D}\partial_{k}^{2} \cr
\equiv\frac{1}{2}\Box_{D}+\frac{1}{2} \sum h^{\mu\nu}({\bf
x})\partial_{\mu}\partial_{\nu}
\end{array}
\end{equation}
and the index at the d'Alambertian denotes its dimensionality. The
inverse in eq.(4) is not unique for the pseudo-Riemannian metric
(2)-(3). However, we define it in the unique way later on by means
of the Feynman proper time representation together with the
$i\epsilon$ prescription.

 Let us begin with the two-point function
(4) and consider the Neumann expansion
\begin{equation}
\begin{array}{l}
\langle (-\Box_{D}-h\partial \partial +M^{2})^{-1}(x,y)\rangle=
\langle\sum_{n=0}^{\infty} ((Gh\partial\partial
)^{n}G)(x,y)\rangle \cr = G(x,y)+
 \partial\partial\partial\partial G G G(x,y)\langle hh\rangle
 +\partial\partial\partial\partial\partial\partial\partial
\partial GGGGG(x,y) \langle hhhh\rangle+...
\end{array}
\end{equation}
where $G=(-\Box_{D}+M^{2})^{-1} $ denotes the Feynman propagator
for the scalar field of mass $M$. In eq.(6) we have used an
integration by parts in order to move the derivatives to the
beginning (this is possible because the correlations of $h$ depend
only on $R^{d}$ and the derivatives concern the complementary
$D-d$ variables). For example, the first non-trivial term in
eq.(6) reads
\begin{equation}
 \int dz_{1}dz_{2}\partial^{X}_{\mu}\partial^{X}_{\nu}\partial_{\sigma}^{X}
 \partial_{\rho}^{X}G(x-z_{1})G(z_{1}-z_{2})G(z_{2}-y)
 {\cal D}^{\mu\nu;\sigma\rho}({\bf z}_{1}-{\bf z}_{2})
\end{equation}
where
\begin{displaymath}
\langle h^{\mu\nu}({\bf z}_{1})h^{\sigma\rho}({\bf z}_{2})\rangle
={\cal D}^{\mu\nu;\sigma\rho}({\bf z}_{1}-{\bf z}_{2})
\end{displaymath}
In order to compare the conventional expansion with the stochastic
one of ref.\cite{jmp} we apply the proper-time representation for
$G$
\begin{equation}
\prod_{k=1}^{n+1}G(v_{k})=\frac{1}{2}\prod_{k=1}^{n+1}\int_{0}^{\infty}id\tau_{k}
\exp(-\frac{i}{2}\tau_{k} M^{2}) p(\tau_{k},{\bf
v}_{k})p(\tau_{k}, {\bf V}_{k})
\end{equation}
where  in the dimension $r$ the Feynman kernel $p$ is
\begin{displaymath}
p(\tau,{\bf x})=(2\pi i\tau)^{-\frac{r}{2}}\exp(-\frac{{\bf
x}^{2}}{2i\tau})
\end{displaymath}
and we made use of the property of the Feynman kernel in
$D$-dimensions that it is a product of the kernels in $D-d$ and
$d$-dimensions. Then, in the product (8) we change the integration
variables, so that
\begin{equation}
\begin{array}{l}
G(x-z_{n})G(z_{n}-z_{n-1}).....G(z_{1}-y) = i^{n+1}2^{-n-1}
\int_{0}^{\infty}d\tau\int_{0}^{\tau}ds_{n}
\int_{0}^{s_{n}}.....\int_{0}^{s_{2}}ds_{1} \cr p(\tau-s_{n},
x-z_{n})p(s_{n}-s_{n-1},z_{n}-z_{n-1})..... p(s_{1},z_{1}-y)
\end{array}
\end{equation}
where in eqs.(8)-(9) we introduced new variables
\begin{displaymath}
\tau_{1}=s_{1}
\end{displaymath}
\begin{displaymath}
\tau_{1}+\tau_{2}=s_{2}
\end{displaymath}
.........
\begin{displaymath}
\tau_{1}+.....+\tau_{n}=s_{n}
\end{displaymath}
\begin{displaymath}
\tau_{1}+.....+\tau_{n+1}=\tau
\end{displaymath}
We write
\begin{displaymath}
{\bf z}_{k}={\bf x}_{k}+{\bf y}
\end{displaymath}
Then, in the representation (9) we apply many times the
Smoluchowski-Kolmogorov equation (the semigroup composition law)
for the $D-d$ dimensional kernels
\begin{equation}
\int p(s-s^{\prime},{\bf X}-{\bf Z})p(s^{\prime},{\bf Z})d {\bf
Z}=p(s,{\bf X})
\end{equation}
Performing the integration by parts we can see that the
derivatives in the expansion (6) (with the representation (9))
 finally act
on $p(\tau,{\bf X}-{\bf Y})$. We may write the result in the
momentum representation in the form (which will come out in the
stochastic version of the next section)
\begin{displaymath}
\partial_{\mu}^{X}p(\tau,{\bf X}-{\bf Y})=
\int d{\bf P}\exp(i{\bf P}({\bf Y}-{\bf
X}))\exp(-i\frac{\tau}{2}{\bf P}^{2} -i\frac{\tau}{2}M^{2} )(-iP_{\mu})
\end{displaymath} We  can compute  higher order correlation functions
 in the Gaussian model of scalar fields. They are expressed by the two-point
 function. For example, the four-point function  expanded
 around the flat metric  reads
\begin{equation}
\begin{array}{l}
\langle
\phi(x)\phi(x^{\prime})\overline{\phi}(y)\overline{\phi}(y^{\prime})\rangle
=\langle( {\cal A}+\frac{1}{2}M^{2})^{-1}\left(x,y\right)( {\cal
A}+\frac{1}{2}M^{2})^{-1}\left(x^{\prime},y^{\prime}\right)\rangle
+(x\rightarrow x^{\prime})
 \cr

 =\langle\sum_{n=0,m=0}^{\infty}
((Gh\partial\partial )^{n}G)(x,y) ((Gh\partial\partial
)^{m}G)(x^{\prime},y^{\prime})\rangle + (x\rightarrow x^{\prime})
\end{array}
\end{equation}
where $x\rightarrow x^{\prime}$ means the same expression with the
exchanged coordinates. In the next section we sum up the
perturbation series (4) and (11). This will allow us to determine
the functional dependence of the scalar correlation functions on the metric
explicitly.

Let us note that the expectation value over the metric field in
eqs.(4) and (11) depends on the
 scalar determinant. It can involve a complex dependence
 on the metric. We cannot compute the effective
 gravitational action explicitly. However,
its  scaling behavior is sufficient to determine the short
distance behavior of the scalar fields.

\section{The stochastic representation}
        Let us recall a
representation of the correlation functions (6) and (11) by means
of the Feynman integral \cite{jmp}\cite{hababook}. In this paper
we work with a real time. It makes no difference for a
perturbation expansion whether we work in Minkowski or in
Euclidean space. However, the non-perturbative Euclidean version
may not exist (e.g. this is the case when the Hamiltonian is
unbounded from below). For this reason we consider here a
representation with an indefinite metric alternative to the
Euclidean version of ref.\cite{jmp}.

 We represent the scalar field two-point function by means of the proper
time method
\begin{equation}
({\cal
A}+\frac{1}{2}M^{2})^{-1}(x,y)=i\int_{0}^{\infty}d\tau
\exp(-\frac{i}{2}M^{2}\tau)\left(\exp\left(-i\tau {\cal A}\right)
\right)(x,y)
\end{equation}
The kernel    $\left(\exp\left(-i\tau {\cal A}\right) \right)(x,y)
$ can be expressed by the Feynman integral
\begin{equation}
\begin{array}{l}
K_{\tau}(x,y)\equiv\left(\exp\left(-i\tau {\cal A}\right)
\right)(x,y)=\int {\cal D}x\exp(\frac{i}{2}\int \frac{d{\bf
x}}{dt}
  \frac{d{\bf x}}{dt}+\frac{i}{2}\int (h^{\mu\nu}({\bf x})+\eta^{\mu\nu})\frac{dX_{\mu}}{dt}
  \frac{dX_{\nu}}{dt})
 \cr
 \delta\left(x\left(0\right)-x\right)
  \delta\left(x\left(\tau\right)-y\right)
  \end{array}
  \end{equation}
In the Feynman integral (13) we make a change of variables ($x
\rightarrow b$) defined by the solution of the Stratonovitch
stochastic differential equations \cite{ike}
\begin{equation}
dx^{\Omega}(s)=e_{A}^{\Omega}\left(
x\left(s\right)\right)db^{A}(s)
\end{equation}
where for $\Omega=1,....,D-d$
\begin{displaymath}
e^{\mu}_{a}e^{\nu}_{a}=\eta^{\mu\nu}+h^{\mu\nu}
\end{displaymath}
and $e^{\Omega}_{A}=\delta^{\Omega}_{A}$ if $\Omega>D-d$.

The change of variables (14) transforms the functional integral
(13) into a Gaussian integral with the covariance
\begin{displaymath}
E[b_{A}(t)b_{C}(s)]=i\delta_{AC}\min(s,t)
\end{displaymath}
  It can be interpreted as an analytic
continuation of the standard Wiener integral.
This means that on a formal level
\begin{displaymath}
E[F]=\int {\cal D}b\exp\Big(\frac{i}{2}\int ds(\frac{db}{ds}
)^{2}\Big)F(b)\end{displaymath} The solution $q_{\tau}$ of eq.(14)
is defined by two vectors $({\bf Q},{\bf q})$ where
\begin{equation}
{\bf q}(\tau,{\bf x})={\bf x}+ {\bf b}(\tau)
\end{equation}
and ${\bf Q}$ has the components (for $\mu=1,...,D-d$)
\begin{equation}
Q^{\mu}(\tau,{\bf X})=X^{\mu}+\int_{0}^{\tau}
e_{a}^{\mu}\left({\bf q}\left(s,{\bf x}\right)\right)dB^{a}(s)
\end{equation}
The kernel (13) can be expressed by the solution (15)-(16)
\begin{displaymath}
\begin{array}{l}
K_{\tau}(x,y)=E[\delta(y-q_{\tau}(x))] \cr = E[\delta({\bf y}
-{\bf x}-{\bf b}(\tau)) \prod_{\mu}\delta\left(Y^{\mu}-Q^{\mu}
\left(\tau,{\bf X}\right)\right)]
\end{array}
\end{displaymath}
When the $\delta$-function is defined by its Fourier transform
then  the kernel $K_{\tau}$ takes the form
\begin{equation}
\begin{array}{l}
K_{\tau}(x,y)=(2\pi)^{-D+d}\int d{\bf P}
 \exp\left(i{\bf P}\left({\bf
Y}-{\bf X}\right)\right)
 \cr E[\delta\left(
{\bf y}-{\bf x}-{\bf
b}\left(\tau\right)\right)
\exp\left(-i\int P_{\mu}e^{\mu}_{a}\left ({\bf
q}\left(s,{\bf x}\right)\right)dB^{a}\left(s\right)\right)]
\end{array}
\end{equation}
We can perform the $B$-integral. The random variables ${\bf b}$
and $B^{a}$ are independent. We can use the formula \cite{ike}
\begin{displaymath}
E[\exp i\int f_{a}({\bf b})dB^{a}]= E[\exp(-\frac{i}{2}\int
f_{a}f_{a}ds)]
\end{displaymath}
Then, the Feynman kernel is expressed by the metric tensor
\begin{equation}
\begin{array}{l}
K_{\tau}(x,y)=(2\pi)^{-D+d}\int d{\bf P}
         \exp\left(i{\bf P}\left({\bf Y}-{\bf
X}\right)\right)\exp(-i\frac{\tau}{2}{\bf P}^{2})
 \cr E[\delta\left(
{\bf y}-{\bf x}-{\bf b}\left(\tau\right)\right)
\exp\left(-\frac{i}{2}\int_{0}^{\tau} P_{\mu}h^{\mu\nu}\left
({\bf q}\left(s,{\bf x}\right)\right)P_{\nu}ds\right)]
\end{array}
\end{equation}
Note that in eq.(18) instead of Feynman paths we may use the
Brownian paths $\tilde{b}$ \cite{hababook} (the $\delta$-function
can be taken away by replacing the Brownian motion by the Brownian
bridge as in \cite{jmp})
\begin{equation}
\tilde{b}=\exp(-i\frac{\pi}{4})b
\end{equation}
Then, the expectation value really is an average over the random
process.

The perturbation expansion in $h$ reads
\begin{equation}
\begin{array}{l}
\langle ({\cal A}+\frac{1}{2}M^{2})^{-1}(x,y)\rangle=
i\int_{0}^{\infty}d\tau
\exp(-\frac{i\tau}{2}M^{2})K_{\tau}(x,y)\cr =(2\pi)^{-D+d}\int
d{\bf P}
         \exp\left(i{\bf P}\left({\bf Y}-{\bf
X}\right)\right) i\int_{0}^{\infty}d\tau
\exp(-\frac{i\tau}{2}{\bf P}^{2}-\frac{i\tau}{2}M^{2})
 E[\delta\left(
{\bf y}-{\bf x}-{\bf b}\left(\tau\right)\right) \cr
\sum_{n=0}^{\infty}(2i)^{-n}\int_{0}^{\tau}ds_{n}\int_{0}^{s_{n}}d
s_{n-1}......\int_{0}^{s_{2}}ds_{1} P_{\mu}h^{\mu\nu}\left ({\bf
q}\left(s_{n},{\bf x}\right)\right)P_{\nu} \cr P_{\sigma}
h^{\sigma\rho}\left ({\bf q}\left(s_{n-1},{\bf
x}\right)\right)P_{\rho}... P_{\alpha}h^{\alpha\beta}\left ({\bf
q}\left(s_{1},{\bf x}\right)\right)P_{\beta}]\end{array}
\end{equation}
The expectation value over the Feynman paths (the complex
 Brownian motion) can be computed and
we obtain as a result the expectation value depending on the
correlation functions of the metric field
\begin{equation}
\begin{array}{l}
\langle\int_{0}^{\infty}d\tau
\exp(-\frac{i\tau}{2}M^{2})K_{\tau}(x,y)\rangle =(2\pi)^{-D+d}\int
d{\bf P}
         \exp\left(i{\bf P}\left({\bf Y}-{\bf
X}\right)\right) i \int_{0}^{\infty}d\tau
\cr
\exp(-\frac{i\tau}{2}{\bf P}^{2}-\frac{i\tau}{2}M^{2})
\sum_{n=0}^{\infty}\int d{\bf x}_{1}....d{\bf x}_{n}d{\bf z}\cr
(2i)^{-n}\int_{0}^{\tau}ds_{n}\int_{0}^{s_{n}}d
s_{n-1}......\int_{0}^{s_{2}} ds_{1}\cr p(s_{1},{\bf
x}_{1})p(s_{2}-s_{1},{\bf x}_{2}-{\bf x}_{1})..... \cr
p(s_{n}-s_{n-1},{\bf x}_{n}-{\bf x}_{n-1})p(\tau-s_{n},{\bf
z}-{\bf x}_{n}) \delta({\bf y}-{\bf x}-{\bf z})\langle
P_{\mu}h^{\mu\nu}\left ({\bf x}+{\bf x}_{1}\right)P_{\nu} \cr ....
P_{\alpha}h^{\alpha\beta}\left ({\bf x}+{\bf
x}_{n}\right)P_{\beta}\rangle\end{array}
\end{equation}
It is easy to see that this formula coincides with eq.(6) where
the representation (9) is applied with the simplification
concerning  the $X$-dependence ( that the result is expressed
finally by $p(\tau,{\bf X}-{\bf Y}$)) discussed at the end of
sec.2.

 Representing the four-point  function by means of the
stochastic method we obtain
 \begin{equation}
\begin{array}{l}
\langle
\phi(x)\phi(x^{\prime})\overline{\phi}(y)\overline{\phi}(y^{\prime})\rangle
= -(2\pi)^{-2D+2d}\int d\tau_{1}d\tau_{2} \int d{\bf P}d{\bf
P}^{\prime}
  \cr
  \exp(-\frac{i}{2}\tau_{1}{\bf P}^{2}-\frac{i}{2}\tau_{2}{\bf P}^{\prime 2}
  -\frac{i}{2}\tau_{1}M^{2}-\frac{i}{2}\tau_{2}M^{ 2})
  \cr
  \exp\left(i{\bf
P}\left({\bf Y}-{\bf X}\right)
 +i{\bf P}^{\prime}\left({\bf Y}^{\prime}-{\bf X}^{\prime}\right) \right)
 \cr E[
\delta\left({\bf y}-{\bf x}-{\bf b}\left(\tau_{1}\right)\right)
\delta\left({\bf y}^{\prime}-{\bf x}^{\prime}-{\bf b}^{\prime}
\left(\tau_{2}\right)\right)
\cr
\exp\left(-\frac{i}{2}\int_{0}^{\tau_{1}} P_{\mu}h^{\mu\nu}\left
({\bf q}\left(s,{\bf x}\right)\right)P_{\nu}ds\right)

\exp\left(-\frac{i}{2}\int_{0}^{\tau_{2}} P^{\prime}_{\mu}h^{\mu\nu}\left
({\bf q}^{\prime}\left(s^{\prime},{\bf x}^{\prime}\right)\right)P^{\prime}_{\nu}
ds^{\prime}\right)
]
\cr
+ (x\rightarrow x^{\prime})
\end{array}
\end{equation}
We make a perturbation expansion in the metric
\begin{equation}
\begin{array}{l}
\sum_{n=0}^{\infty}(2i)^{-n}\int_{0}^{\tau_{1}}ds_{n}\int_{0}^{s_{n}}d
s_{n-1}......\cr ...\int_{0}^{s_{2}}ds_{1} P_{\mu}h^{\mu\nu}\left
({\bf q}\left(s_{n},{\bf x}\right)\right)P_{\nu} \cr P_{\sigma}
h^{\sigma\rho}\left ({\bf q}\left(s_{n-1},{\bf
x}\right)\right)P_{\rho}.... P_{\alpha}h^{\alpha\beta}\left ({\bf
q}\left(s_{1},{\bf x}\right)\right)P_{\beta} \cr
\sum_{m=0}^{\infty}(2i)^{-m}\int_{0}^{\tau_{2}}ds^{\prime}_{m}
\int_{0}^{s^{\prime}_{m}}d
s^{\prime}_{m-1}......\int_{0}^{s^{\prime}_{2}}ds^{\prime}_{1}
P^{\prime}_{\mu}h^{\mu\nu}\left ({\bf
q}^{\prime}\left(s^{\prime}_{m},{\bf
x}\right)\right)P^{\prime}_{\nu} \cr P^{\prime}_{\sigma}
h^{\sigma\rho}\left ({\bf q}^{\prime}\left(s^{\prime}_{m-1},{\bf
x}^{\prime}\right)\right) P^{\prime}_{\rho}....
P^{\prime}_{\alpha}h^{\alpha\beta}\left ({\bf
q}^{\prime}\left(s^{\prime}_{1},{\bf
x}\right)\right)P^{\prime}_{\beta} + (x\rightarrow x^{\prime})
\end{array}
\end{equation}
The paths ${\bf q}$ and ${\bf q}^{\prime}$ are independent. We
calculate the expectation values according to the formula (21). In
the conventional perturbation expansion (11) we make the same
change of the proper time variables (9) as in the two-point
function (for each two-point function separately). Then, it
becomes evident by the same argument as the one at the two-point
function that the stochastic formula (23) and the conventional one
(11) coincide (no matter what are the correlation functions for
the metric field). It is now easy to see that the argument
concerning the four-point function can be generalized to any
$2n$-point function.
\section{Ultraviolet behavior of the perturbation series}
In order to study the ultraviolet behavior of the perturbation
expansion (6) and (11) it is useful to express the functional
integral in the momentum space. First, we write down the scalar
part of the action (1) in the form
\begin{equation}
\begin{array}{l}
L=\int d{\bf x}d{\bf P}(\nabla_{\bf x}\overline{\phi}({\bf x},{\bf
P}) \nabla_{\bf x}\phi({\bf x},{\bf P}) +{\bf
P}^{2}\overline{\phi}({\bf x},{\bf P}) \phi({\bf x},{\bf P})\cr
+M^{2}\overline{\phi}({\bf x},{\bf P}) \phi({\bf x},{\bf P})
+h^{\mu\nu}({\bf x})P_{\mu}P_{\nu}\overline{\phi}({\bf x},{\bf P})
\phi({\bf x},{\bf P}))
\end{array}
\end{equation}
Then, it is easy to see from the functional integral
for the correlation functions
\begin{displaymath}
\int {\cal D}\phi\exp(-L)\overline{\phi}....\phi
\end{displaymath}
that there is a direct correspondence between the model (1) and
the trilinear interaction
$P_{\mu}P_{\nu}h^{\mu\nu}\overline{\phi}\phi$ where $P$ is treated
just as a parameter.
 The Fourier transform over $P$ is performed in the final formula. For example
\begin{equation}
\begin{array}{l}
\langle \overline{\phi}({\bf x},{\bf X})\phi({\bf x}^{\prime},{\bf
X}^{\prime})\rangle \cr =\int d{\bf P}d{\bf P}^{\prime}\langle
\overline{\phi}({\bf x},{\bf P}) \phi({\bf x}^{\prime},{\bf
P}^{\prime})\rangle
        \exp(i{\bf PX}-i{\bf P}^{\prime}{\bf X}^{\prime})
\cr =\int d{\bf P}\langle(-\triangle_{d}+{\bf P}^{2}+h^{\mu\nu}
P_{\mu}P_{\nu} )^{-1}({\bf x},{\bf x}^{\prime})\rangle
        \exp(i{\bf P}({\bf X}-{\bf X}^{\prime}))
\end{array}
\end{equation}
whereas for the four-point function
\begin{equation}
\begin{array}{l}
\langle \overline{\phi}({\bf x},{\bf X})\overline{\phi}({\bf
x}^{\prime},{\bf X}^{\prime}) \phi({\bf y},{\bf Y}) \phi({\bf
y}^{\prime},{\bf Y}^{\prime})\rangle \cr

=\int d{\bf P}d{\bf P}^{\prime}d{\bf K}d{\bf K}^{\prime} \langle
\overline{\phi}({\bf x},{\bf P})\overline{\phi}({\bf
x}^{\prime},{\bf P}^{\prime}) \phi({\bf y},{\bf K}) \phi({\bf
y}^{\prime},{\bf K}^{\prime})\rangle \cr
        \exp(i{\bf PX}+i{\bf P}^{\prime}{\bf X}^{\prime}-
        i{\bf KY}-i{\bf K}^{\prime}{\bf Y}^{\prime})
\cr =\int d{\bf P}d{\bf K}\langle(-\triangle_{d}+{\bf
P}^{2}+h^{\mu\nu} P_{\mu}P_{\nu} )^{-1}({\bf x},{\bf y})
 \cr
      (-\triangle_{d}+{\bf K}^{2}+h^{\mu\nu}
K_{\mu}K_{\nu} )^{-1}({\bf x}^{\prime},{\bf y}^{\prime})\rangle
        \exp(i{\bf P}({\bf X}-{\bf Y}))
            \exp(i{\bf K}({\bf X}^{\prime}-{\bf Y}^{\prime}))
      \cr
        + (x\rightarrow x^{\prime}  )

\end{array}
\end{equation}
 The integration over the $D-d$ momenta  P does not
 lead to any additional infinities.
 It follows that the divergencies depend on the dimension d
and on the singularity of the metric field correlations. Hence, if
$d<4$ and $h$ has the canonical short distance behavior $\vert
{\bf x}-{\bf y}\vert^{-d+2}$ then there will be no divergencies at
all.

 \section{Modified short distance behavior }

 The correlation functions (12) and (22) have a different scaling behavior
 in ${\bf x}$ and ${\bf X}$ directions. We  obtain
 a fixed scaling behavior either
  setting ${\bf x}    ={\bf y}=0    $ or ${\bf X}={\bf Y}=0$.
Assume that $h^{\mu\nu}$ is scale invariant at short distances
\begin{equation}
h^{\mu\nu}({\bf x})\simeq \lambda^{2\gamma}h^{\mu\nu}(\lambda{\bf
x})
\end{equation}
Then,  by scaling (just as in \cite{jmp}, although
$g^{\mu\nu}=\eta^{\mu\nu} + h^{\mu\nu}$ is not scale invariant)
and using the stochastic formula (18) we obtain
  \begin{equation}
  \langle {\cal A}^{-1}(x,y)\rangle\simeq  \vert
  {\bf X}-{\bf Y}\vert^{-D+2-\frac{\gamma}{1-\gamma}(d-2)}
  \end{equation}
  at short distances.
   This argument
   can be extended to all  correlation functions
   showing that
  \begin{equation}
  \phi(0,{\bf X})\simeq
  \lambda^{\frac{D-2}{2}+\frac{\gamma}{1-\gamma}\frac{d-2}{2}}
  \phi(0,\lambda{\bf X})
\end{equation}
where the equivalence means that both sides have the same
correlation functions at short distances.

    For the behavior in the ${\bf x}$ direction we let  $X=Y=0$.
In such a case just by scaling  of momenta in eq.(18) we can bring
the propagator  to the form
\begin{equation}
\langle {\cal A}^{-1}(x,y)\rangle=\int_{0}^{\infty}d\tau
\tau^{-\frac{d}{2}-(D-d)(1-\gamma)/2}F_{2}(\tau^ {-\frac{1}{2}}({\bf
y}    -{\bf x}))
\end{equation}
 Hence
   \begin{equation}
  \langle {\cal A}^{-1}({\bf x},{\bf y})\rangle=  R  \vert
  {\bf x}-{\bf y}\vert^{-d+2-(D-d)(1-\gamma)}
  \end{equation}
  where $R$ is a  constant.
   In general one can show similarly as in eq.(29) that
  \begin{equation}
  \phi({\bf x},0)\simeq
  \lambda^{\frac{d-2}{2}+\frac{(D-d)(1-\gamma)}{2}}
  \phi(\lambda{\bf x},0)
\end{equation}

We could see that the results (29) and (32) do not require the
stochastic representation (18). For the two-point function we
write the expansion (6) in the (equivalent) form (21). Using the
scale invariance (27) and changing the integration variables we
can rewrite the perturbation series (21) in the form
\begin{equation}
\begin{array}{l}
\langle {\cal A}^{-1}(x,y)\rangle = \lambda^{-2+d+(1-\gamma)(D-d)}
\int_{0}^{\infty}d\tau\int d{\bf P}\exp(i{\bf P}\lambda^{1-\gamma}
({\bf X}-{\bf Y})) \cr \exp(-i\tau \lambda^{-2\gamma}({\bf
P}^{2}+M^{2}))\sum_{n}f_{n}(\tau,{\bf P},\lambda({\bf y}-{\bf x}))
\end{array}
\end{equation}
We set $\lambda=\vert {\bf y}-{\bf x}\vert^{-1}$ if ${\bf X}={\bf
Y}$ and $\lambda=\vert {\bf X}-{\bf Y}\vert^{-\frac{1}{1-\gamma}}
$ if ${\bf x}={\bf y}$. The formulas (28) and (31) follow under
the assumption that setting ${\bf P}^{2}+M^{2}=0$ in eq.(33) we
obtain a finite result from the integral over $\tau$ of the sum of
$f_{n}$.
  In a similar way we can obtain
 the short distance behavior of higher order correlation functions.
As an example we could consider a dimensional reduction of the
gravitational action (a static approximation) from $d=4$ to $d=3$.
Then, in the quadratic approximation to the threedimensional
quantum gravity we obtain $\gamma=\frac{1}{4}$ and an explicit
finite average over $h$ in eq.(18).
\section{Discussion}
We have obtained formulas expressing quantum scalar field
correlation functions by quantum gravitational correlation
functions in a model where the metric field depends only on a
lower dimensional submanifold. Such models can either be
considered as an approximation to a realistic theory or they may
come from  a dimensional reduction of a higher dimensional
Einstein gravity \cite{cr}- \cite{nicolai}. In either case the
non-perturbative phenomena are essential for the short distance
behavior. In this way we have confirmed by a resummation of the
perturbation series our results \cite{PLB} based on a formal
functional integral. The result is that a singular quantum
gravitational field substantially modifies the short distance
behavior of the matter fields. In particular, it can make the
matter fields more regular.

\end{document}